\documentclass[aps,prl,twocolumn,showpacs,groupedaddress,letterpaper,fleqn]{revtex4}  
\usepackage{graphicx}  
\usepackage{dcolumn}   
\usepackage{bm}        
\usepackage{amssymb}   
\usepackage{fleqn}

\newcommand{\etal}{\emph{et al.}}
\newcommand{\ks}{\ensuremath{K_S^0}}
\newcommand{\lb}{\ensuremath{\Lambda_b^0}}
\newcommand{\lc}{\ensuremath{\Lambda_c^+}}

\begin{document}


\hspace{5.2in} \mbox{Fermilab-Pub-07/205-E}

\title{Measurement of the $\lb$ lifetime using semileptonic decays}
%
\author{V.M.~Abazov$^{35}$}
\author{B.~Abbott$^{75}$}
\author{M.~Abolins$^{65}$}
\author{B.S.~Acharya$^{28}$}
\author{M.~Adams$^{51}$}
\author{T.~Adams$^{49}$}
\author{E.~Aguilo$^{5}$}
\author{S.H.~Ahn$^{30}$}
\author{M.~Ahsan$^{59}$}
\author{G.D.~Alexeev$^{35}$}
% ...

\affiliation{\vspace{0.1 in}(The D\O\ Collaboration)\vspace{0.1 in}}
% ...
\affiliation{$^{5}$University of Alberta, Edmonton, Alberta, Canada,
                Simon Fraser University, Burnaby, British Columbia, Canada,
                York University, Toronto, Ontario, Canada, and
                McGill University, Montreal, Quebec, Canada}
% ...
\affiliation{$^{28}$Tata Institute of Fundamental Research, Mumbai, India}
% ...
\affiliation{$^{30}$Korea Detector Laboratory, Korea University, Seoul, Korea}
% ...
\affiliation{$^{35}$Joint Institute for Nuclear Research, Dubna, Russia}
% ...
\affiliation{$^{49}$Florida State University, Tallahassee, Florida 32306, USA}
% ...
\affiliation{$^{51}$University of Illinois at Chicago,
                Chicago, Illinois 60607, USA}
% ...
\affiliation{$^{59}$Kansas State University, Manhattan, Kansas 66506, USA}
% ...
\affiliation{$^{65}$Michigan State University,
                East Lansing, Michigan 48824, USA}
% ...
\affiliation{$^{75}$University of Oklahoma, Norman, Oklahoma 73019, USA}
% ...
\date{June 15, 2007}

\begin{abstract}
  We report a measurement of the $\lb$ lifetime
  using a sample corresponding to 1.3~fb$^{-1}$ of
  data collected by the D0\ experiment in 2002--2006
  during Run~II of the Fermilab Tevatron collider. The $\lb$ baryon
  is reconstructed via the decay $\lb \to  \mu \bar \nu\lc  X$.
  Using $4437 \pm 329$ signal candidates, we measure the $\lb$ lifetime to be
  $\tau(\lb)$ = 1.290 ${}^{+0.119}_{-0.110}$ (stat) $^{+0.087}_{-0.091}$ (syst) ps,
  which is among the most precise measurements in semileptonic $\lb$ decays.
  This result is in good agreement with the world average value.
\end{abstract}

\pacs{14.20.Mr, 14.40.Nd, 13.30.Eg, 13.25.Hw}

\maketitle 

Lifetimes of $b$ hadrons provide an important test of models
describing quark interaction within bound states.
The experimental measurement of the lifetimes are in
reasonable agreement with the theoretical 
predictions~\cite{theory,tarantino,gabbiani}, 
but further improvement in the experimental and theoretical 
precision is essential for the development of non-perturbative 
quantum chromodynamics. 

The lifetime of $b$ baryons recently attracted a special
interest. The current world average $\lb$ lifetime is 
$\tau(\lb) = 1.230 \pm 0.074$ ps, and the
ratio of the $\lb$ baryon and $B^0$ meson lifetimes is 
$\tau(\lb) / \tau(B^0) = 0.80 \pm 0.05$~\cite{pdg},
in good agreement with the theoretical prediction 
$\tau(\lb) / \tau(B^0) = 0.86 \pm 0.05$ \cite{gabbiani}. 
However, the recent $\lb$ lifetime measurement from the CDF collaboration 
in the $\lb \to J/\psi \Lambda$ decay gives a significantly larger
value: $\tau(\lb) = 1.593 ^{+0.083} _{- 0.078} \pm 0.033$ ps~\cite{cdf}, 
not included in the quoted world average.
Additional $\lb$ lifetime measurements could provide a potential resolution
of this inconsistency.

This Letter presents a measurement of the $\lb$ lifetime
using the semileptonic decay $\lb \to \mu \bar{\nu} \lc X$,
where $X$ is any other particle.  
Charge conjugated states are implied throughout this paper. The $\lc$ baryon
is selected in the decay $\lc \rightarrow \ks p$.
The sample corresponds to approximately 1.3 fb$^{-1}$ 
of data collected by the D0\ experiment in Run II
of the Fermilab Tevatron Collider.

The D0\ detector is described in detail elsewhere~\cite{run2det}. The
components most important to this analysis are the central
tracking and muon systems. The central tracking system consists
of a silicon microstrip tracker and a central fiber tracker,
both located within a 2 T superconducting solenoidal magnet,
with designs optimized for tracking and vertexing at pseudorapidities
$|\eta|<3$ and $|\eta|<2.5$ respectively (where $\eta$ =
$-$ln[tan($\theta$/2)] and $\theta$  is the polar angle of the particle with respect 
to the proton beam direction).  The muon system is located outside the
calorimeters and has pseudorapidity coverage $|\eta|<2$.  It consists
of a layer of tracking detectors and scintillation trigger counters in
front of 1.8 T iron toroids, followed by two similar layers after the
toroids~\cite{run2muon}. The trigger system identifies events of interest in 
a high-luminosity environment based on muon identification and charged
tracking. Some triggers require a large impact parameter for the muon.
Since this condition biases the lifetime measurement, the events selected
exclusively by these triggers are removed from our sample.
All processes and decays required for this analysis are simulated using the 
{\sc evtgen} \cite{EvtGen} generator interfaced 
to {\sc pythia} \cite{pyth} and followed 
by full modeling of the detector response using {\sc geant} \cite{geant} 
and event reconstruction.

Reconstruction of the $\lb$ decay starts from the selection of a muon,
which must have 
at least two track segments in the muon chambers associated with a central 
track, with transverse momentum $p_T > 2.0$ GeV/$c$.
All charged particles in the event are clustered into jets using 
the Durham clustering 
algorithm~\cite{DURHAM}. The products of the $\lc$ decay 
are then searched for among tracks belonging to the jet 
containing the identified muon.

The primary vertex is determined using the method 
described in Ref.~\cite{PV}.
The $\ks$ meson is reconstructed as a combination of two 
oppositely charged tracks that have a common vertex
displaced from the $p \bar p$ interaction point by at least 
four standard deviations of the measured decay length in the plane
perpendicular to the beam direction. 
Both tracks are assigned the pion mass
and the mass of the $\pi^+ \pi^-$ system is required to be
consistent with the $\ks$ mass to within 1.8 standard deviations. 
Combinations consistent with the $\Lambda \to p \pi$ hypothesis, 
when either track is assigned the proton mass and the
mass of the $p \pi$ system lies between 1.109 and 1.120 GeV/$c^2$, are rejected.
Any other charged track in the jet with $p_T > 1.0$ GeV/$c$ 
and at least two hits in the silicon detector is assigned
the proton mass and combined with  
the neutral extrapolated $\ks$ candidate to form a $\lc$ candidate.
Their common vertex is required to have a fit $\chi^2$/d.o.f.$< 9/1$.
The $\lc$ candidate is combined with 
the muon to make a $\lb$ candidate, and its invariant mass 
is required to be between 3.4 and 5.4 GeV/$c^2$.
A common vertex for the $\lc$ candidate and muon is required to have
a fit $\chi^2$/d.o.f. $< 9/1$. 
The transverse distance $d_T^{bc}$ between the $\lb$ 
and $\lc$ vertices is calculated and is assigned a positive sign
if the $\lb$ vertex is closer to the primary vertex, and a negative
sign otherwise. The $\lb$ candidate is required to have 
$-3.0 < d_T^{bc}/\sigma(d_T^{bc}) < 3.3$, where $\sigma(d_T^{bc})$ 
is the uncertainty of the $d_T^{bc}$ measurement. The upper bound on 
the distance between $\lb$ and $\lc$ vertices reduces
the background significantly,
since the $\lc$ lifetime is known to be very small: 
$0.200 \pm 0.006$ ps~\cite{pdg}.

To further improve the $\lb$ signal selection, 
a likelihood ratio method \cite{bgv}
is utilized. This method provides a simple way to combine many 
discriminating variables into a single variable with an increased power
to separate signal and background. The variables chosen for this analysis
are the $\lb$ isolation, the transverse momentum of the $\ks$, proton
and $\lc$ candidates, and the mass of the $\mu \lc$ system.
The isolation is defined as the fraction 
of the total momentum of charged particles 
within a cone around the $\mu \lc$ direction
carried by the $\lb$ candidate. The cone is defined by the condition
$\sqrt{(\Delta\eta)^2 + (\Delta\phi)^2} < 0.5$, where 
$\Delta\eta$ and $\Delta\phi$ are the difference in 
pseudorapidity and azimuthal angle 
from the direction of the $\lb$ candidate.

\begin{figure}[tbh]
\begin{center}
\includegraphics[width=8cm]
{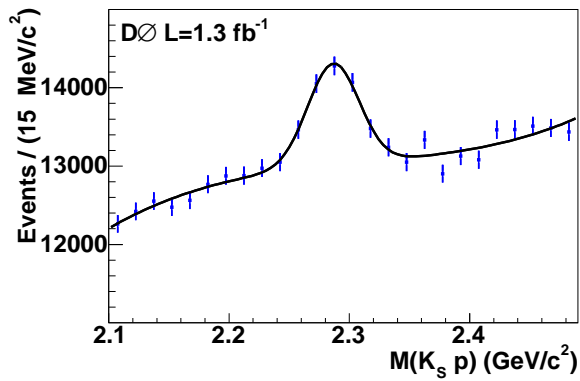}
\caption[]{The $\ks p$ invariant mass for the selected $\lb$ 
candidates and fit overlaid (see text). 
Notice the suppressed-zero scale of the vertical axis.} 
\label{fig1}
\end{center}
\end{figure}

Figure \ref{fig1} shows the invariant  mass $M(\ks p)$ for 
the selected $\lb$ candidates.
The fit to this distribution is performed 
with a signal Gaussian function and a fourth-order polynomial 
function for the background.  
The  $\lc$ signal contains 4437 $\pm$ 329 (stat) events 
at a central mass of 2285.8 $\pm$ 1.7 MeV/$c^2$. 
The width of the mass peak is $\sigma = 20.6 \pm 1.7$ MeV/$c^2$ consistent
with that observed in the simulation.

Simulation shows that the contribution from the
$B_d \to \ks \pi$ decay when a pion is assigned the proton mass 
has a broad $M(\ks p)$ distribution with no excess in 
the $\lc$ mass region.

Since the final state is not fully reconstructed,
the $\lb$ proper decay length cannot be determined. 
Instead, a measured visible proper decay length $\lambda^M$
is computed as $\lambda^M = $ $m c \left( \bm{L}_T \cdot
\bm{p}_T(\mu \lc) \right) / 
|\bm{p}_T(\mu \lc)|^2$. $\bm{L}_T$
is the vector from the primary vertex to the $\lb$ 
vertex in the plane perpendicular to the beams, 
$\bm{p}_T(\mu \lc)$ is the transverse momentum of
the $\mu \lc$ system and $m = 5.624$ GeV/$c^2$ 
is taken as the $\lb$ mass~\cite{pdg}. 

To determine the $\lb$ lifetime, 
the selected sample is split into a number of $\lambda^M$ bins.
The mass distribution in each bin is fitted with 
a signal Gaussian and a fourth degree polynomial background.
The position and width of the Gaussian are fixed to the values obtained
from the fit of the entire sample (see Fig.~\ref{fig1}). 
The Gaussian normalization
and background parameters are allowed to float in the fit.
The range of $\lambda^M$ and the number of signal events fitted 
in each bin $n_i$ together with its statistical uncertainty $\sigma_i$ 
are shown in Table \ref{tab2}.  

\begin{table}[tbh]
\caption{Fitted signal yield in different $\lambda^M$ bins}
\begin{ruledtabular}
\newcolumntype{A}{D{A}{\pm}{-1}}
\begin{tabular}{cA}
$\lambda^M$ range(cm)  & 
   \multicolumn{1}{c}{Number of signal candidates $n_i \pm \sigma_i$~(stat)} \\
\hline 
 $[-0.06,-0.04]$ & 62\, A \,48 \\
 $[-0.04, -0.02]$ & 66\, A \,69 \\
 $[-0.02, 0.00]$ & 587\, A \,156 \\
 $[ 0.00 , 0.02]$ & 1172\,A \,173 \\
 $[ 0.02 , 0.04]$ & 999\,A \,99 \\
 $[ 0.04 , 0.06]$ & 540\,A \,69 \\
 $[ 0.06 , 0.08]$ & 299\,A \,54 \\
 $[ 0.08 , 0.10]$ & 225\,A \,44 \\
 $[ 0.10 , 0.20]$ & 454\,A \,64 \\
 $[ 0.20 , 0.30]$ & 47\,A \,34 \\
\end{tabular}
\end{ruledtabular}
\label{tab2}
\end{table}

The expected number of signal events in each bin $n_i^e$ is given by
$n_i^e = N_{tot} \int_{i} f(\lambda^M)d\lambda^M$, 
where $N_{tot}$ is the total number of $\mu \lc$ events, 
and $f(\lambda^M)$ is the probability density function ({\it pdf}) 
for $\lambda^M$. The integration is done within the range of a given bin. 

In addition to $\lb \to \mu \bar \nu \lc X$ decays, the 
$\lc$ baryon can also be created in $c \bar{c}$ or $b \bar b$
production, along with a muon from the decay of the second $c$ or $b$ hadron. 
In what follows, these processes are referred to as {\it peaking background},
since they produce a $\lc$ peak in the $\ks p$ mass spectrum imitating
the signal.
Such events are reconstructed as $\lb$ candidates, and have 
a fake vertex formed by the intersection of the muon and $\lc$ trajectories.
The simulation shows that the distribution of $\lambda^M$
for such a fake vertex has a mean of zero and 
a standard deviation of $\approx$150 $\mu$m. 

The expression for $f(\lambda^M)$ takes into account the contribution of 
signal and peaking background:
$f(\lambda^M) = (1-r_{\rm bck})f_{\rm sig}(\lambda^M) + r_{\rm bck} f_{\rm bck}(\lambda^M)$. 
Here $r_{\rm bck}$ is the fraction of peaking background, and 
$f_{\rm sig}(\lambda^M)$ and $f_{\rm bck}(\lambda^M)$ are the {\it pdf}'s 
for signal and background respectively. 
The background {\it pdf} is taken from the simulation.
The signal {\it pdf} is expressed as the convolution of the decay probability
and the detector resolution:
$f_{\rm sig}(\lambda^M) = \int dK H(K) \left[ \theta(\lambda) K/(c\tau) 
\exp(-K\lambda/(c\tau)) \otimes R(\lambda^M-\lambda,s)\right]$.
Here, $\tau$ is the $\lb$ lifetime, and $\theta(\lambda)$ is the step function.
The factor $K = p_T(\mu \lc)/p_T(\lb)$ is a measure of
the difference between the measured $p_T(\mu \lc)$ and true momentum of the $\lb$ candidate,
and $H(K)$ is its {\it pdf}. The $R(\lambda^M-\lambda,s)$ 
is a function modeling the detector resolution.
A scale factor $s$ accounts for the difference between the
expected and actual $\lambda^M$ resolution.

The $H(K)$ distribution is obtained from the simulation. 
The contribution of decays $\lb \to \mu \bar \nu \lc $ and
$\lb \to \mu \bar \nu \Sigma_c \pi$ with $\Sigma_c \to \lc \pi$
is taken into account. 
The contributions of $\lb \to \lc D_s^{(*)-}$ with the $D_s^-$ decaying 
semileptonically, $\Xi_b \to \mu \bar \nu \Lambda_c X$ and 
$\lb \to \tau^- \bar \nu \lc$ with $\tau^- \to \mu^- \bar \nu_{\mu} \nu_{\tau}$
are found to be strongly suppressed by
the branching fractions and low reconstruction efficiency. 
To obtain $H(K)$, the $K$ factor distribution of each process
is weighted with its expected fraction in the selected sample.
This is computed taking into account both the
reconstruction efficiency and the branching fraction of each process.
The fraction of $ \ell^- \bar{\nu} \lc$ in semileptonic
$\lb$ decays has been measured recently to be 
$0.47 {}^{+0.12}_{-0.10}$  \cite{pdg}.
We use this result in our analysis. 

The resolution function is given by
$R(\lambda^M-\lambda,s) = \int f_{\rm res}(\sigma) 
G(\lambda^M-\lambda,\sigma,s)d\sigma$,
where $f_{\rm res}(\sigma)$ is the {\it pdf} for the expected 
resolution of $\lambda^M$,
and $G$ is a Gaussian function
$G(\lambda^M-\lambda,\sigma,s) = 1/ (\sqrt{2\pi}\sigma s) 
\exp[-(\lambda_M-\lambda)^2/ (2 \sigma^2 s^2)].$ The $\sigma_s$ is 
the decay length uncertainty, which is
determined for each candidate from the track parameter uncertainties
propagated to the vertex uncertainties.

To determine $f_{\rm res}(\sigma)$,
signal and background subsamples are defined
according to the mass of the $\ks p$ system. 
All events with 
$2244.7 < M(\ks p) < 2326.9$ MeV/$c^2$ are included in the signal
subsample, and all events with  $2183.9 < M(\ks p) < 2225.0$ MeV/$c^2$ and 
$2346.6 < M(\ks p) < 2387.7$ MeV/$c^2$ are included in the background 
subsample. In addition, the events in both subsamples are required
to have a measured proper decay length exceeding 200 $\mu$m. This cut
reduces the background under the $\lc$ signal 
and the contribution of peaking background.
The  $f_{\rm res}(\sigma)$ distribution is obtained by subtracting
the distribution of expected resolution in the background subsample
from the distribution in the signal subsample.

The $\lb$ lifetime is determined by the minimization of 
$\chi^2 = \sum_i^{N_{bins}} (n_i - n_i^e)^2/\sigma_i^2$, where the sum is taken
over all bins of measured proper decay length (Table~\ref{tab2}).
The free parameters of the fit are $N_{tot}$, $\tau(\lb)$ and 
$r_{\rm bck}$. A separate study is performed to 
measure the resolution scale factor
using the decay $D^{*+} \to D^0 \pi^+$ with
$D^0 \to \mu^+ \nu \ks \pi^-$. It has a similar topology to that of the 
$\lb \to \mu \bar \nu \lc$ decay. Since the $D^{*+}$ meson comes mainly from 
$c \bar c$ production, its decay vertex coincides with the primary interaction point.
The distribution of the $D^{*+}$ proper decay length is mainly
determined by the detector resolution and can be used to measure
the resolution scale factor. A value of $1.19 \pm 0.06$ is found.
The scale factor in the lifetime fit is fixed to this value 
and varied later in a wide range to estimate an associated 
systematic uncertainty.

\begin{figure}
\begin{center}
\includegraphics[width=8cm]{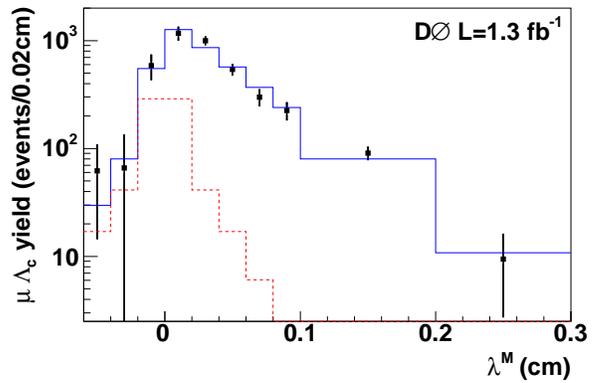}
\caption{Measured $\mu \lc$ yields in the $\lambda^M$ bins (points) 
and the result of 
the lifetime fit (solid histogram).  The dashed histogram shows 
the contribution of peaking background.}
\label{fig5}
\end{center}
\end{figure}

The lifetime fit gives 
$\tau (\lb) = 1.290^{+0.119}_{-0.110}$ (stat) ps,
and the fraction of peaking background
$r_{\rm bck} = 0.160^{+0.068}_{-0.074}$ (stat). 
Figure \ref{fig5} shows the distribution of the number of $\lc \mu$
events versus $\lambda^M$ together 
with the result of the lifetime fit superimposed.
The lifetime model agrees well with data with a $\chi^2$/d.o.f.= 5.5/7.
The dashed line shows separately the contribution of the peaking background.

The method used to fit the mass distribution in each of the $\lambda^M$ bins
is the most significant source of systematic uncertainty.
The fit sensitivity is tested by refitting each $\lambda^M$ bin for
the mass interval between 2.17 and 2.40 GeV/$c^2$ with a linear
parametrization of the background. Binning effects of the mass histograms
are checked by performing fits to the data with bins of half the nominal width
and with the lowest and highest bins excluded.
The lifetime fit is performed 
again for each test.  The largest deviation of $\tau(\lb)$ 
is 0.067 ps, which is given as the systematic uncertainty due to the 
mass-fitting procedure.
The parameters describing the peaking background are varied
by their uncertainties.
The largest shift in the fitted $\lb$ lifetime is 0.012 ps.  

The selected sample can also contain a contribution from
$B \to \mu \bar \nu \lc X$ decay. Its branching fraction is unknown;
only the upper limit Br$(B \to e \bar \nu \lc X) < 3.2 \times 10^{-3}$ at 90\% CL 
is available \cite{pdg}. The possible contamination from this decay 
would reduce the fitted $\lb$ lifetime, since the $K$ factor for these
events is smaller. The upper 90\% CL limit on the fraction of this decay in the
selected sample is estimated to be 5\%, which would result in the reduction
of the $\lb$ lifetime by 0.027 ps. 

The value of the scale factor is varied by $\pm 20\%$, and shifts 
of approximately $\pm 0.036$ ps are observed in the fitted lifetime.
This value is also included in the systematic uncertainty.

The fraction of $\lb \to \mu \bar \nu \lc$ decay in the semileptonic $\lb$ decays
is varied between 0.3 and 0.6. The lower bound is selected to be larger
than the current uncertainty in this fraction \cite{pdg} to take into
account the possible contribution from 
decays to $ \tau \bar \nu \lc$ and 
other heavier states with lower mean $K$ factor.  
The shift of 0.025 ps 
in the fitted lifetime is taken as the systematic uncertainty due to 
the branching fractions in the $K$ factor.   
The mean of the $K$ factor distribution does not change significantly with 
the $p_T$ of the muon, however the shape of the distribution is changed.  
To estimate the possible variation of the $\lb$ lifetime, 
the distribution for $\mu \bar \nu \lc $ decays 
is generated with a cut of $p_T(\mu) > 6$ GeV/$c$ and the fit is 
repeated.
A shift of 0.005 ps is observed, which is assumed as the uncertainty
due to the momentum dependence of the $K$ factor.
 The change in the $K$ factor distribution due to the uncertainty in generation
and decay of $B$ hadrons has been estimated in other analyzes 
to be less than 2\% \cite{bpb0,burd}.  Therefore we shift all $K$ factor 
values by $\pm 2\%$, and observe a shift of 0.026 ps in the 
fitted lifetime.  
The overall systematic uncertainty due to the $K$ factor distribution
is estimated to be 0.036 ps.
The effect on lifetime measurement due to misalignment of elements of the tracking detector 
is determined by rescaling the geometrical position of all detectors within uncertainties
of the alignment procedure. The resulting variation of the $\lb$
lifetime is estimated to be 0.018 ps.

The systematic uncertainties are summarized and added in quadrature in Table 
\ref{tab3}. Total systematic uncertainty of this measurement
is estimated to be 0.09 ps.

In addition, several consistency checks of this analysis are
performed. The fitting procedure is applied to the simulated 
$\lb \to \mu \bar \nu \lc$ events that passed 
the full reconstruction chain and all selection criteria
used in data. The fitted lifetime is consistent with the generated value.
The simulated events are also used to test that the measured
proper decay length is not biased with respect to the generated 
one, and that
the applied selections have the same efficiency for different values
of $\lb$ lifetime.

To test for any bias produced by the fitting 
procedure, 500 fast, parameterized Monte Carlo samples are generated
and analyzed.
The average lifetime 
agrees with the generated one, and the assigned uncertainty corresponds
to the statistical spread of fitted values.

Another test consists of splitting the data
sample into two roughly equal parts using various criteria 
and measuring the $\lb$ lifetime in each sample independently.
The sample is split according to the muon charge, the muon direction,
the decay length of $\ks$
or the chronological date of data taking. All such tests give statistically
consistent values of the $\lb$ lifetime.

In conclusion, our measurement of the $\lb$ lifetime 
using the semileptonic decay $\lb \to \mu \bar \nu \lc X$ results in
$\tau(\lb)  =  1.290^{+0.119}_{-0.110}~\mbox{(stat)}^{+0.087}_{-0.091}
~\mbox{(syst) ps.}$
It is consistent with the current world average $\lb$ lifetime
and with our measurement 
in the exclusive decay $\lb \to J/\psi \Lambda$ \cite{lbjpsil}.
The D\O\ results are statistically independent and the correlation
of systematics between them is very small. Their combination results in
$\tau(\lb) = 1.251^{+0.102}_{-0.096}$~ps.
Our new measurements are less consistent with the recent discrepant
measured $\lb$ lifetime \cite{cdf} than with the current world average \cite {pdg}.


\begin{table}[t]
\caption{\label{tab3}Systematic uncertainties in $\tau(\lb)$}
\begin{ruledtabular}
\begin{tabular}{cc}
Source & Uncertainty in $\tau(\lb)$\\
\hline
Detector alignment & $\pm 0.018$ ps\\
Mass-fitting method & $\pm 0.067$ ps\\
$K$-factor determination & $\pm 0.036$ ps\\
Peaking background & $\pm 0.012$ ps\\
Resolution scale factor & $\pm 0.036$ ps\\
Contribution of $B \to \mu \bar \nu \Lambda_c X$ & $^{+0.000}_{-0.027}$ ps 
\vspace*{0.1cm} \\

\hline
\vspace*{-0.3cm} & \\
Total & $^{+0.087}_{-0.091}$ ps\\
\end{tabular}
\end{ruledtabular}
\end{table}

%
We thank the staffs at Fermilab and collaborating institutions, 
and acknowledge support from the 
DOE and NSF (USA);
CEA and CNRS/IN2P3 (France);
FASI, Rosatom and RFBR (Russia);
CAPES, CNPq, FAPERJ, FAPESP and FUNDUNESP (Brazil);
DAE and DST (India);
Colciencias (Colombia);
CONACyT (Mexico);
KRF and KOSEF (Korea);
CONICET and UBACyT (Argentina);
FOM (The Netherlands);
Science and Technology Facilities Council (United Kingdom);
MSMT and GACR (Czech Republic);
CRC Program, CFI, NSERC and WestGrid Project (Canada);
BMBF and DFG (Germany);
SFI (Ireland);
The Swedish Research Council (Sweden);
CAS and CNSF (China);
Alexander von Humboldt Foundation;
and the Marie Curie Program.
%

\end{document}